# HIGH-FREQUENCY DIELECTRIC SPECTROSCOPY OF $BaTiO_3$ CORE - SILICA SHELL NANOCOMPOSITES: PROBLEM OF INTERDIFFUSION

D. NUZHNYY, J. PETZELT, V. BOVTUN, M. KEMPA, M. SAVINOV
*Institute of Physics ASCR, Na Slovance 2, 18221 Praha 8, Czech Republic*

C. ELISSALDE, U-C. CHUNG, D. MICHAU, M. MAGLIONE
*CNRS, Université Bordeaux, ICMCB, F-33608 Pessac Cedex, France*

C. ESTOURNÈS
*CIRIMAT et Plateforme Nationale CNRS de Frittage Flash, PNF2 MHT,*
*Université Paul Sabatier, 33062 Toulouse, France*

Three types of $BaTiO_3$ core – amorphous nano-shell composite ceramics were processed from the same core-shell powder by standard sintering, spark-plasma sintering and two-step sintering techniques and characterized by XRD, HRSEM and broad-band dielectric spectroscopy in the frequency range $10^3$ - $10^{13}$ Hz including the THz and IR range. The samples differed by porosity and by the amount of interdiffusion from the cores to shells, in correlation with their increasing porosity. The dielectric spectra were also calculated using suitable models based on effective medium approximation. The measurements revealed a strong dielectric dispersion below the THz range, which cannot be explained by the modeling, and whose strength was in correlation with the degree of interdiffusion. We assigned it to an effect of the interdiffusion layers, giving rise to a strong interfacial polarization. It appears that the high-frequency dielectric spectroscopy is an extremely sensitive tool for detection of any gradient layers and sample inhomogeneities even in dielectric materials with negligible conductivity.

*Keywords*: core-shell nanocomposites, infrared and THz spectroscopy, barium titanate, effective medium approximation.

## 1. Introduction

Recently, a lot of activity has been devoted to process and study the core-shell composite and nanocomposite ceramics with strongly differing electric and/or magnetic properties of cores and shells. The basic idea is to prepare materials with new properties not available in homogeneous systems. For instance, in the case of ferroelectric – ferromagnetic core-shell systems the idea is to introduce an effective magnetoelectric coupling absent in individual components [1-5]. But new properties can be expected even if the microscopic coupling of both components is negligible and both components preserve their bulk properties, since compact shells prevent any percolation of the core properties to macroscopic distances. The most striking effect in this respect is the possibility to change conducting core materials into dielectrics or even giant effective permittivity dielectrics in ceramic-metal composites (cermets) [6] or barrier-layer capacitors [7]. Even in dielectric systems with negligible conductivity, the effect of shells can dramatically influence the effective dielectric properties. For instance, in high-permittivity (e.g. ferroelectric) ceramics, even if they are single phase, the grain boundaries usually display a much lower local permittivity (so called dead layers) and play a role of the shells, which can dramatically reduce the effective permittivity leading to a pronounced dielectric size effect in nano-grain ceramics [8].

However, in addition to down scaling of the grain size (which requires various techniques of nano-powders synthesis), recent evolution of the bulk electroceramics for passive components applications requires also interface control including grain boundary engineering and polycrystalline multi materials assembly. In the case of ferroelectric-based composites a combination of ferroelectric (barium titanate BT or barium-strontium titanate BST) and low-permittivity dielectric (MgO, $MgTiO_3$, $SiO_2$, $Al_2O_3$, $BaZrO_3$) components is challenging to reduce the dielectric losses. In this context, there is an obvious interest to propose novel multi-materials assembly and to find a subtle compromise between the composition, architecture and micro/nanostructure to reach the desired properties. Interdiffusion at interfaces, distribution and scale of inhomogeneities (second phases, defects, etc.…) and nano/microstructure are parameters which, if they are not well identified, can obscure the properties of the composites. These key parameters emphasize the intimate link between the thermal treatment and the aimed macroscopic properties in these bulk electroceramics. The success of the "multi-materials route" for tailoring properties depends strongly on the complex problem of both interfaces and micro/nanostructure control.

Recently, some of us have studied high-frequency dielectric properties of dense BT@$BaZrO_3$ and BT@$SrTiO_3$ core–shell composites prepared by Spark Plasma Sintering (SPS) [9]. A strong terahertz (THz) and microwave (MW) absorption and dispersion were observed, which were not present in any of the pure components. Modeling of the dielectric response of such composites by effective medium approach (EMA), which assumes sharp boundaries between the components, predicts smaller THz and MW losses in the composite than in any of the components. The observed strongly enhanced absorption was assigned to a small interdiffusion of BT cores into the shells, observed also by electron microscopy [10]. This means that such interdiffusion creates a rather strong interfacial polarization in the gradient layers between the cores and shells. Particularly strong effects could be expected if the cores possess some free charge carriers whose diffusion could contribute to this interfacial polarization. This could lead to supercapacitor effect with effective low-frequency permittivity values in the range of $10^5$, as recently observed in reduced-BT@$SiO_2$ composites [11,12]. Moreover, the ferroelectric phase transition, which appears in homogeneous solid solutions of the same integral composition, vanishes or is completely smeared out in such composites. It appears that the high-frequency dielectric spectroscopy is extremely sensitive to such interdiffusion effects.

Similar additional MW and THz losses were also observed in our BT@alumina composites [13], where also a small interdiffusion was detected [14]. In this paper we report on the THz and IR spectroscopy study of several BT - silica (BT@$SiO_2$) composites with the aim to prove the correlation between the additional high-frequency losses and interdiffusion. Different nanoscale functionalisation of the ferroelectric grains by a dielectric shell can be used to tailor the dielectric losses of such a composite. The soft chemistry route we have used to prepare the ferroelectric core@shell grains as starting building blocks, ensure a uniform, continuous and homogeneous nanoscale coating (from 1 to 100 nm) of individual BT particles [1,15]. The full coverage of each particle is definitely a critical condition to reach a control of the chemical and structural mismatch between the core and shell at the grain scale. Very recently, some of us have reported a high level of interface probing in silica coated BT ceramics [16]. An accurate control of the interphase formation between the two components was possible thanks to an interface study at the atomic scale. A secondary phase, fresnoite $Ba_2TiSi_2O_8$, was identified, growing in-between the two components. By adjusting the thermal treatment it is possible either to favor the fresnoite formation or to preserve the amorphous silica shell in between the ferroelectric grains. In the latter case, the material can be considered as a network of disconnected BT particles with silica shells at grain boundaries.

We propose, starting from initial silica coated BT particles and using different sintering processes (standard, SPS and two-step sintering), to tune the interface at the grain scale and thus to change the micro/nanostructure of the resulting ceramics. We show that IR, THz and MW spectroscopies are efficient probes of the interdiffusion occurring at the core@shell interface, allowing an accurate interpretation of the dielectric properties and interface relationships.

## 2. Experimental

BT particles of the 500 nm mean diameter (BT500) were purchased from Sakai Chemical Co. (Japan). Silica coating (~5 nm) is obtained using a derived method from the so called Stöber process. The surface of the BT particles has to be activated by acidic treatment with nitric and citric acids before the silica coating. The reaction takes place in a water/alcohol/ammonia solution using tetraethoxysilane (TEOS) as the silica source. The procedure was described in details elsewhere [15]. Using different sintering processes, we have prepared three nanostructured ceramics with different values of the relative density according to the different degree of preservation of the initial core@shell design.

Standard sintering (S) was applied on the BT@silica powder at 600°C during 2h with heating rates 100°C/h. The density remains low ~58% and no secondary phase occurs on the XRD pattern (S sample). Nabertherm RHTC 40-450/15 tubular furnace was used for conventional sintering. The microstructural observation of the ceramic fracture was performed using a high resolution scanning electron microscope JEOL 6700F (Fig. 1). Another BT@silica powder was sintered using Spark Plasma Sintering (SPS - Dr Sinter SPS-2080 SPS Syntex INC Japan of the Plateforme Nationale de Frittage Flash (PNF2) of CNRS at Toulouse (France)). The temperature was raised to 600 °C over a period of 3 min, and from this point it was monitored and regulated by an optical pyrometer focussed on a small hole located at the surface of the die. A heating rate of 100 °Cmin$^{-1}$ was used to reach the final temperature of 1000 °C under argon (Ar) atmosphere. The sintering time was 15 minutes and a uniaxial pressure of 100 MPa was applied. The density of the sample was close to 80% (SPS sample). Finally, BT@silica ceramics were also obtained using a two-step sintering at $T_1$= 1225 °C (1 min) and $T_2$= 900 °C (12 h) with two additional steps at 250 and 600°C of one hour each (TSS sample). The heating rates were 100 °C/h except between 600 and 900 °C where it was 500 °C/h. Its density was 95%. The formation of second phase - fresnoite - was clearly identified on the XRD pattern performed at room temperature on the as prepared ceramics (TSS sample). The concentration of fresnoite was estimated from the XRD pattern comparing the diffraction peaks intensity of pure $BaTiO_3$ and fresnoite. In our case fresnoite/BaTiO3 = 0.07**,** which means that almost the whole shells were transformed into fresnoite [16] To study the influence of fresnoite, we processed also a pure fresnoite ceramics. $BaCO_3$, $TiO_2$ and $SiO_2$ were used as precursors for preparation of the fresnoite powder *via* solid state reaction. Fresnoite ceramics was obtained by conventional sintering at 1350 C during 3 hours under oxygen flow. Density of the obtained ceramics was ~ 85%.

Ceramic disks with thicknesses of ~1.3 mm and diameters of ~8 mm and ~5 mm were used for IR reflection measurements. SPS sample was polished and the other samples were used as prepared. For THz transmission measurements the TSS sample was thinned down to ~60 μm and the SPS sample down to ~54 μm.

The IR reflectivity measurements under near-normal incidence were performed using FTIR Bruker IFS 113v spectrometer equipped with pyroelectric deuterated triglycine sulfate detectors in the range of 25 – 700 cm$^{-1}$. A custom-made time-domain THz spectrometer, based on a femtosecond Ti:sapphire laser [17] and using interdigited photoconducting switch for generation of THz pulses and electro-optic sampling scheme with [110] ZnTe crystal as the THz detector, was used to measure the complex transmission from which the complex dielectric response was directly calculated in the range of 5 – 30 cm$^{-1}$ (~ 150 – 900 GHz). For high-temperature (300 – 900 K) IR reflection and THz transmission measurements the samples were placed into commercial high-temperature cell (Specac P/N 5850).

The samples were further studied by HF and MW dielectric measurements (dielectric spectrometer with Novocontrol BDS 2100 coaxial sample cell and Agilent 4291B impedance analyzer in the 1 MHz - 1.8 GHz range, open-end coaxial technique with Agilent E8364B vector network analyzer in the 200 MHz - 8 GHz range and resonance dielectric measurements at 5.8 and 12.3 GHz), and by standard low-frequency dielectric measurements (dielectric analyzer Novocontrol Alpha AN in the $10^{-2}$ – $10^{6}$ Hz range).

The IR reflectivity spectra were fitted with the factorized form of the complex permittivity [18]

$$R(\omega) = \left| \frac{\sqrt{\varepsilon(\omega)} - 1}{\sqrt{\varepsilon(\omega)} + 1} \right|^2 , \qquad \varepsilon(\omega) = \varepsilon_\infty \prod_j \frac{\omega_{LOj}^2 - \omega^2 + i\omega\gamma_{LOj}}{\omega_{TOj}^2 - \omega^2 + i\omega\gamma_{TOj}} \tag{1}$$

where $\omega_{TOj}$ and $\omega_{LOj}$ are the frequencies of the *j*-th transverse optic (TO) and longitudinal optic (LO) polar mode, respectively, $\gamma_{TOj}$ and $\gamma_{LOj}$ are their damping constants, respectively, and $\varepsilon_\infty$ is the optical electronic contribution to the permittivity.

At room temperature, the complex dielectric functions obtained by using (1) were compared with the calculated effective dielectric functions using the generalized brick-wall model [19] based on the EMA

$$\varepsilon_{cs} = \varepsilon_b V_b + (1 - V_b)\frac{\varepsilon_a \varepsilon_b}{(1-n)\varepsilon_b + n\varepsilon_a} \quad (2)$$

where $V_b$ is the volume fraction of the shell component with the zero depolarizing field: $V_b = \frac{x-n}{1-n}$, $0 < n < x$ ($x$ is the volume concentration of the shells) and the depolarizing factor $n = gx$, $0 \le g \le 1$, where the geometrical factor $g$ characterizes the topology of grains (g = 1/3: coated spheres). First, the coated-spheres model was applied for the S and SPS samples to obtain the dielectric response of core-shell system $\varepsilon_{cs}$. We used the room-temperature dielectric function of dense BT nanoceramics with ~50 nm grain size [20] and 5.8 vol% of silica with the constant $\varepsilon_b = 5$. Afterwards, the effective dielectric response of real samples was calculated using the Bruggeman EMA formula [21]

$$(1-x)\frac{\varepsilon_1 - \varepsilon_{eff}}{\varepsilon_1 + 2\varepsilon_{eff}} + x\frac{\varepsilon_2 - \varepsilon_{eff}}{\varepsilon_2 + 2\varepsilon_{eff}} = 0 \quad (3)$$

where $\varepsilon_1$ is the effective dielectric function of the core-shell system and $\varepsilon_2 = 1$ is the permittivity of pores, considering the known porosity $x$ of the samples.

For the TSS sample, effective dielectric response was calculated using the Bruggeman EMA twice, for both mixing the BT cores with silica/fresnoite shell due to the non-complete coverage of the BT cores and for taking into account the porosity of the sample.

### 3. Results, Evaluation, Modeling and Discussion

In Fig. 2 we present the room temperature IR reflectivity spectra of our samples below 700 cm$^{-1}$ (relevant polar phonon range of BT) combined with the low-frequency reflectivity (~6-25 cm$^{-1}$) calculated from our THz data. The fit of the data using Eq. (1) is added by thin solid lines. The rather large differences in the reflectivity levels reflect the different porosity of the samples, but partly are also due to the surface quality (mainly the TSS sample, for which the overall reflectivity level is decreasing with frequency indicating surface scattering effects). Temperature dependences of the THz dielectric data for increasing temperatures (up to 700-900 K) are shown in Fig. 3a,b,c for the sample S, SPS and TSS, respectively. Notice the non-monotonous dependences of both permittivity and loss, showing up maxima between 400-500 K, manifesting the ferroelectric phase transition in BT. Notice also the strong dielectric dispersion towards low frequencies below the THz range in the SPS and TSS samples, which in the S sample is substantially reduced.

The strength of the dielectric dispersion and the temperature dependences are better seen from Fig. 4, where for all samples the temperature dependences of complex permittivity are plotted at fixed frequencies in the kHz range, at 30 MHz and at ~250 GHz. For the S sample, no measurable dielectric dispersion (ε' ~ 18) was seen at room temperature in the whole frequency range 1 MHz – 10 GHz. The ferroelectric transition is revealed in all the samples, but the most pronounced temperature dependence and dispersion is seen for the TSS sample. Note the large permittivity differences between the low and high frequencies in the case of TSS and SPS sample, whereas for the S sample this difference (i.e. dispersion below the polar phonon range) is very small.

In Figs. 5, 6, 7 we compare the fitted reflectivity spectra and calculated dielectric functions from the fit parameters for the S, SPS and TSS sample, respectively. Note the log frequency scale for emphasizing the soft-mode region. For the SPS sample we also measured the temperature dependence on heating from which in Fig. 6 we show the spectra at 600 K. We also plotted the model spectra at room temperature taking the spectra of BT spheres (from BT ceramics with 50-nm grain size [20], see also [9]) coated with silica approximated by a constant permittivity $\varepsilon_b$ = 5 (using Eq. (2) with $g$ = 1/3) and the known porosity (using Eq. (3) with $\varepsilon_2$ = 1). One can see that the qualitative agreement with the spectra from the fit is good (particularly the sharp TO2 mode near 180 $cm^{-1}$ and the reflectivity minima near 470 $cm^{-1}$ as well as the TO4 mode near 480 $cm^{-1}$ are well reproducible by the model, which has no free parameter). However, the strengths of the broad TO1 (soft) mode (averaged over the strongly split E and $A_1$ response in the ferroelectric phase) are reduced in the experimental spectra of the S and TSS sample, but for the SPS sample it looks opposite. In the latter case the TO1 mode is softer than modeled, which indicates imperfect shells (percolation regions of the BT cores). Due to known non-complete coverage of the BT cores in the case of TSS sample we modeled the influence of shells by the Bruggeman model (Eq. (3)) rather than by the coated spheres model. The rather large difference in the effective dielectric response of S sample between the modeled and experimental data can be explained by higher porosity of the sample. A good agreement of modeled spectra with the experiment can be achieved using the S sample porosity of 63% .

Since in case of the SPS sample a fresnoite second phase was revealed in the shell region [16], we also studied the pure fresnoite ceramics to check its dielectric properties, in the low-frequency range already published [22]. The fitted IR reflectivity and calculated dielectric function are shown in Fig. 8. We are not going to discuss in detail the rich polar mode spectrum, but it is clear that the IR contribution to permittivity as well as the directly measured low-frequency permittivity [22] remain below ~12 without appreciable dielectric dispersion below the THz range. Therefore the influence of the second phase is not expected to have an important impact on the effective dielectric properties and we left the modeling of the shell spectra by the constant $\varepsilon_b$ = 5.

Let us now discuss the dielectric dispersion below the polar phonon range, which is evident from Figs. 3 and 4 and was also observed in the previously studied BT@$SrTiO_3$ and BT@$BaZrO_3$ systems [9]. In principle, EMA models, which assume ideally sharp boundaries among the components, cannot yield additional absorption below the lowest-frequency loss peaks of the constituents, because the effect of depolarizing field is just to shift up the absorption regions in the spectra of the softer component (with the lowest-frequency loss peak) [19]. Our dielectric function of BT taken from [20] does not include any dielectric dispersion below the soft mode region in the THz range (see also Fig. 6 in Ref. 9). In fact it neglects the additional dispersion due to possible domain-wall contribution in the GHz range, which is however not very strong in BT ceramics at room temperature [20,23] Moreover, our data show that even above $T_C$, where no ferroelectric domains can contribute, there is a pronounced dielectric dispersion in our SPS and TSS composites in the THz range and below. Therefore the observed additional dispersion should be caused by some gradient layers between the cores and shells or neighbouring core-shell grains. Since the contact among the neighbouring core-shell grains is not connected with any inhomogeneity in case of ideal coverage of individual cores by the shells, the only reason for the dispersion could be some interdiffusion of BT cores into the silica shells. This picture is in correlation with the directly observed interdiffusion, which was the strongest in case of the most dense TSS sample with the highest low-frequency permittivity. It was somewhat weaker in case of the SPS sample and it was almost negligibly small in case of the most porous S sample, where no interdiffusion was observed due to the low sintering temperature.

The reason for the pronounced dispersion due to these gradient layers between the cores and shells has to be some charges (presumably bound) inevitably connected with such interfacial gradient layers, since the ionic charges are not compensated in such inhomogeneous layers, at least in an ideal case without defect charges. Since the gradient layers are quite thin (of the order of 1 nm) the quantitative understanding of such a strong effect calls for some modeling or first-principles calculation.

## 4. Conclusions

Our high-frequency dielectric spectra of the three core-shell BT@$SiO_2$ composites differing in the amount of interdiffusion from the BT cores into the silica shells have revealed an additional THz-MW dispersion whose strength is in correlation with the amount of interdiffusion. Its origin consists presumably in bound charges connected with these interfacial core-shell layers, creating in this way a strong interfacial polarization. It appears that the dielectric spectroscopy is an extremely sensitive tool for detecting any inhomogeneities and gradient layers in ceramics and composite samples. Similar effects, but in much lower frequency range, are well known from ceramics with non-negligible (defect or ionic) conductivity differing in the grain bulk and boundaries (Maxwell-Wagner effect), see e.g. [24]. More detailed measurements in the whole frequency range are needed to try the description of the additional dispersion by some more complicated model similar to doubly-coated spheres suggested for $SrTiO_3$ ceramics [25] or more complex brick-layer models [24].

From this features, it becomes clear that it is extremely difficult to estimate and tailor dielectric losses (as well as permittivity) in a broad frequency range just by mixing low and high loss/permittivity dielectrics in a composite without paying a careful attention to the presence of possible interdiffusion and interfacial layers.

## Acknowledgments

The work was supported by the Academy of Sciences of the Czech Republic (project AVOZ 10100520), the Czech Science Foundation (project 202/09/0430) and by COST action MP904.

## Figure captions

**Fig. 1.** HRSEM picture of the S sample (density 58%).

**Fig. 2.** Room temperature IR reflectivity of the three studied BT@$SiO_2$ composites. The thin solid lines represent the multioscillator fits (Eq. (1)).

**Fig. 3.** Temperature dependent dielectric data from the time-domain THz spectroscopy. Notice the non-monotonous dependences. (a) S sample, (b) SPS sample, (c) TSS sample.

**Fig. 4.** Temperature dependent dielectric data at selected frequencies (indicated in the Figure) from the kHz to THz range. Note the log scale for permittivity $\varepsilon'$.

**Fig. 5.** EMA model (Eqs. (2),(3)) of the room temperature IR dielectric function of the S sample (dashed lines) compared to that obtained from the reflectivity fit (Eq. (1)) and THz data. The remarkable difference between the modeled effective spectra and experimental data is probably due to a higher porosity of the S sample.

**Fig. 6.** EMA model (Eqs. (2),(3)) of the room temperature IR dielectric function of the SPS sample compared to that obtained from the reflectivity fit (Eq. (1)) and THz data at 300 and 600 K. The difference between modeled effective spectra and experimental data indicates a presence of imperfect shells of the BT cores, not taken into account by the coated spheres model (Eq. 2).

**Fig. 7.** EMA model (Eq. (3)) of the IR room temperature dielectric function of the TSS sample (dashed lines) compared to that obtained from the reflectivity fit (Eq. (1)) and THz data. The higher values of the modeled effective spectra compared with the experimental data can be explained by using the Bruggeman model (Eq. 3).

**Fig. 8.** Room temperature IR reflectivity of the fresnoite ceramics (open symbols) and its fit (Eq. (1)) and calculated IR dielectric function (solid lines).

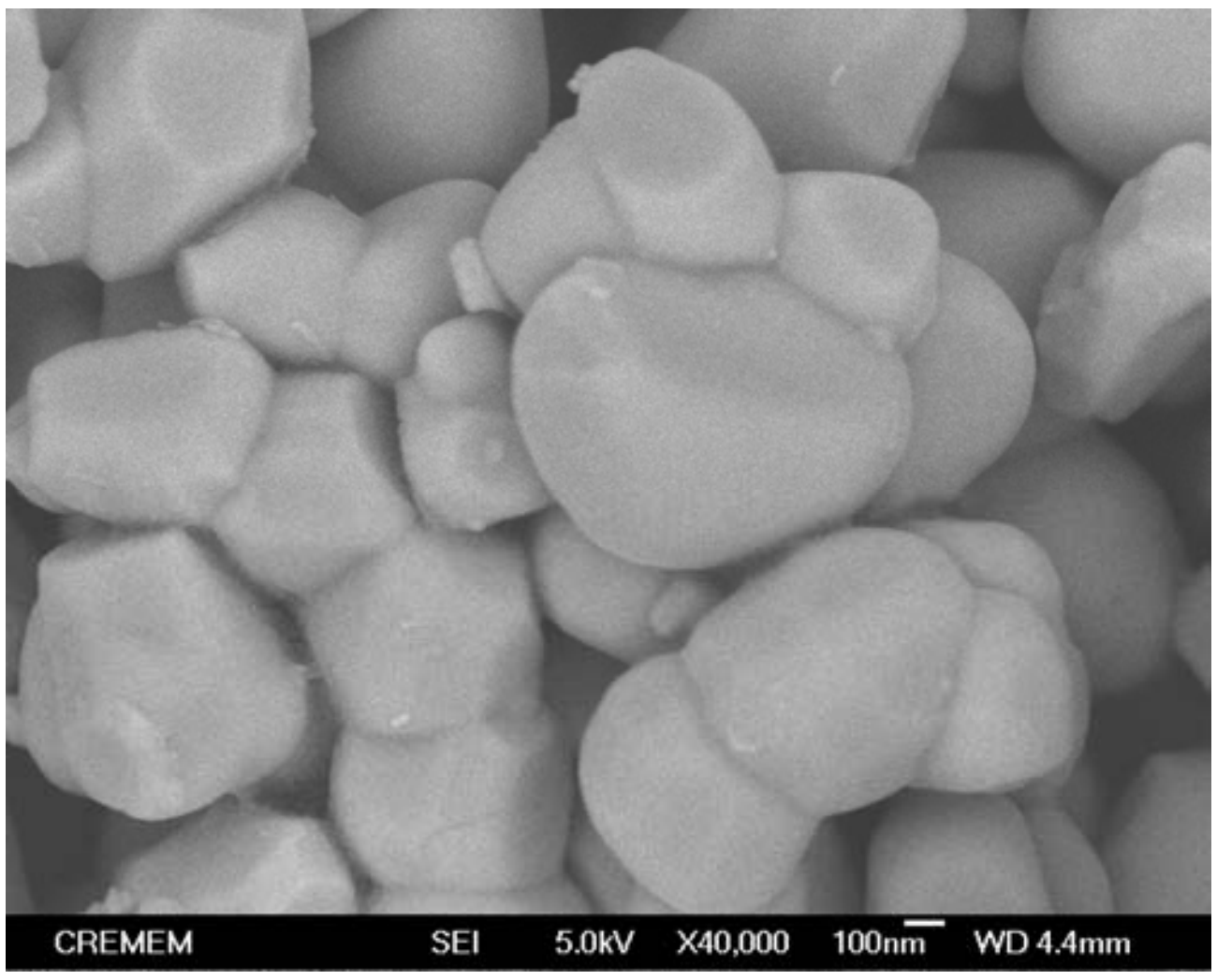

Fig.1.

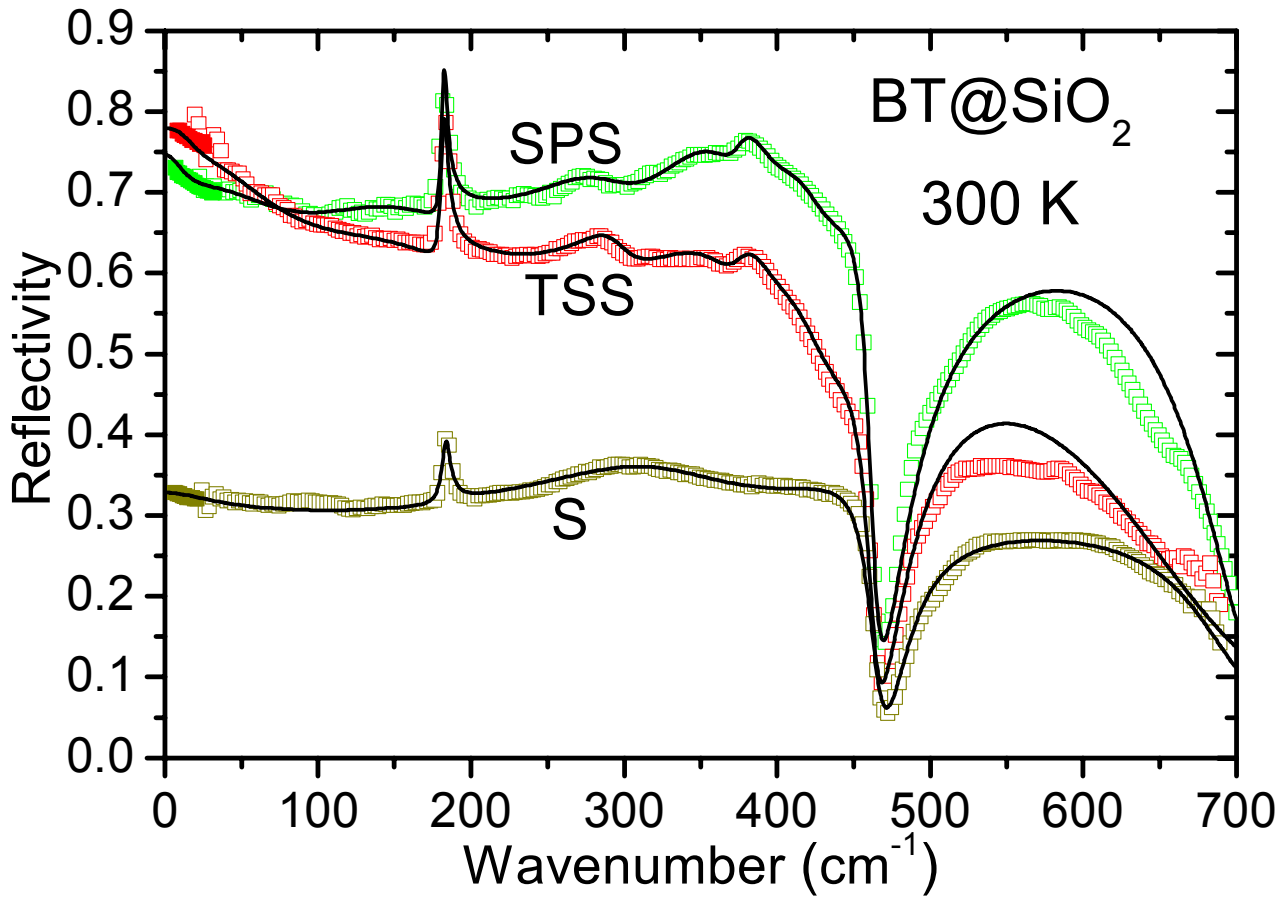

Fig. 2.

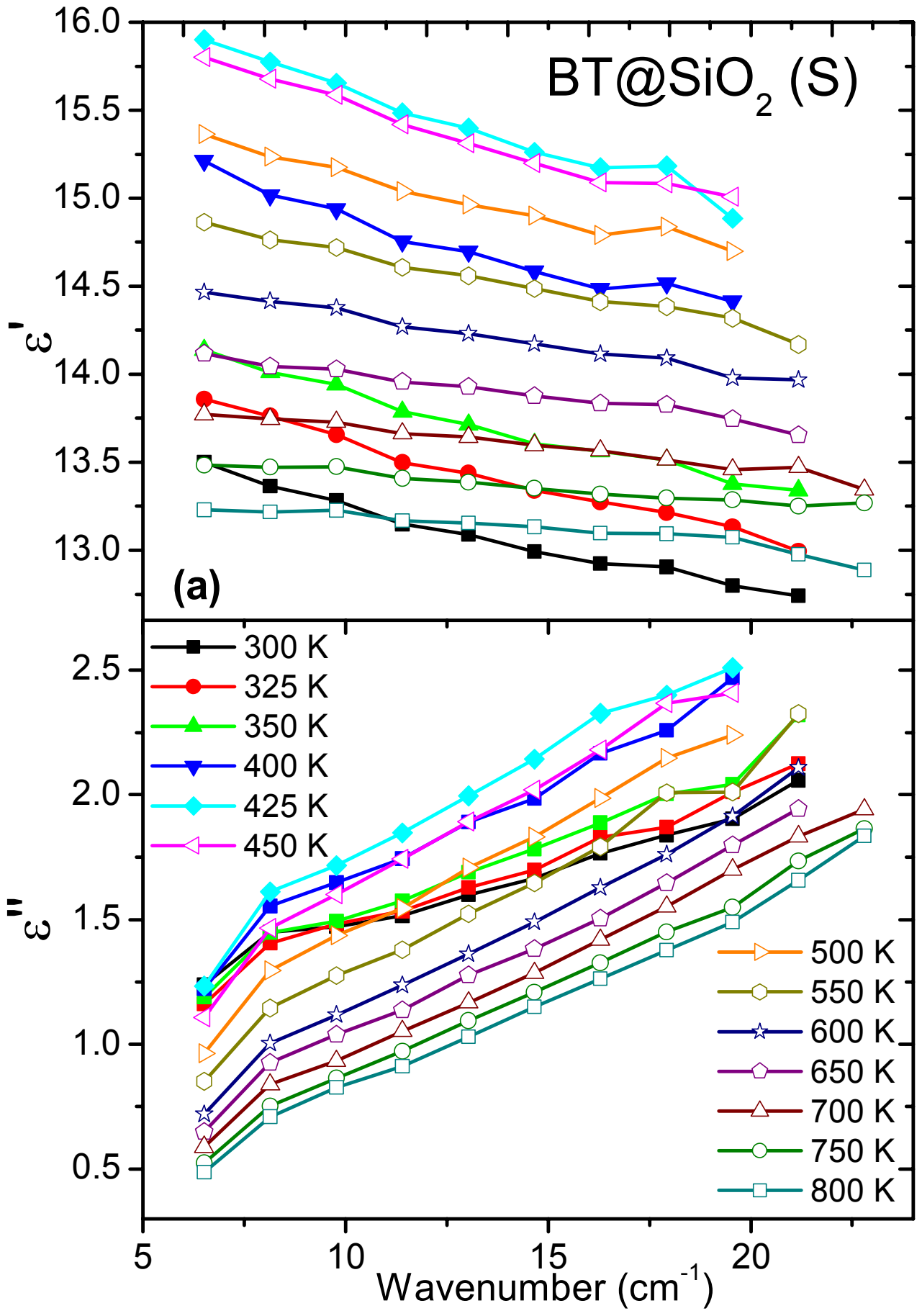

Fig. 3(a).

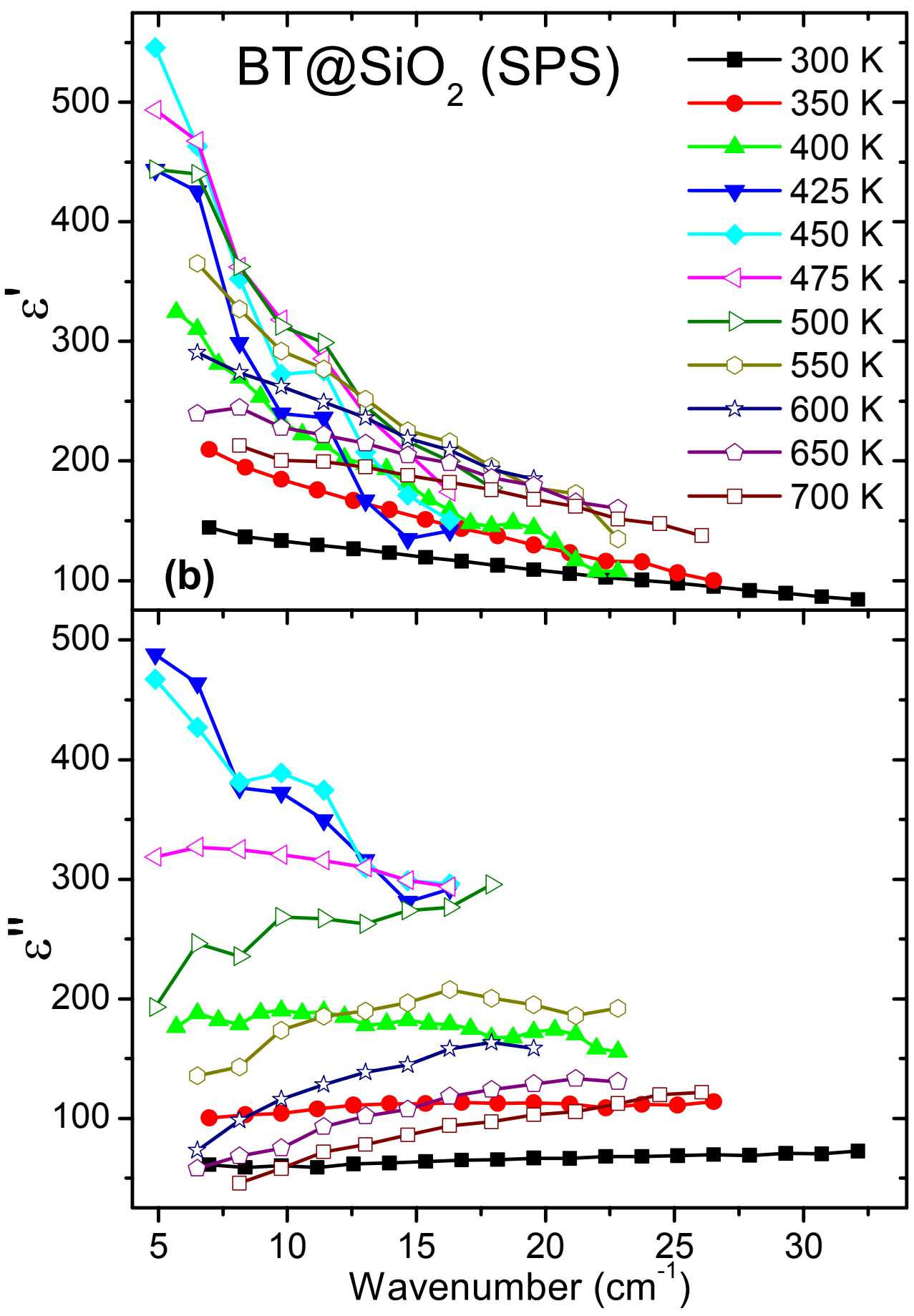

Fig. 3(b).

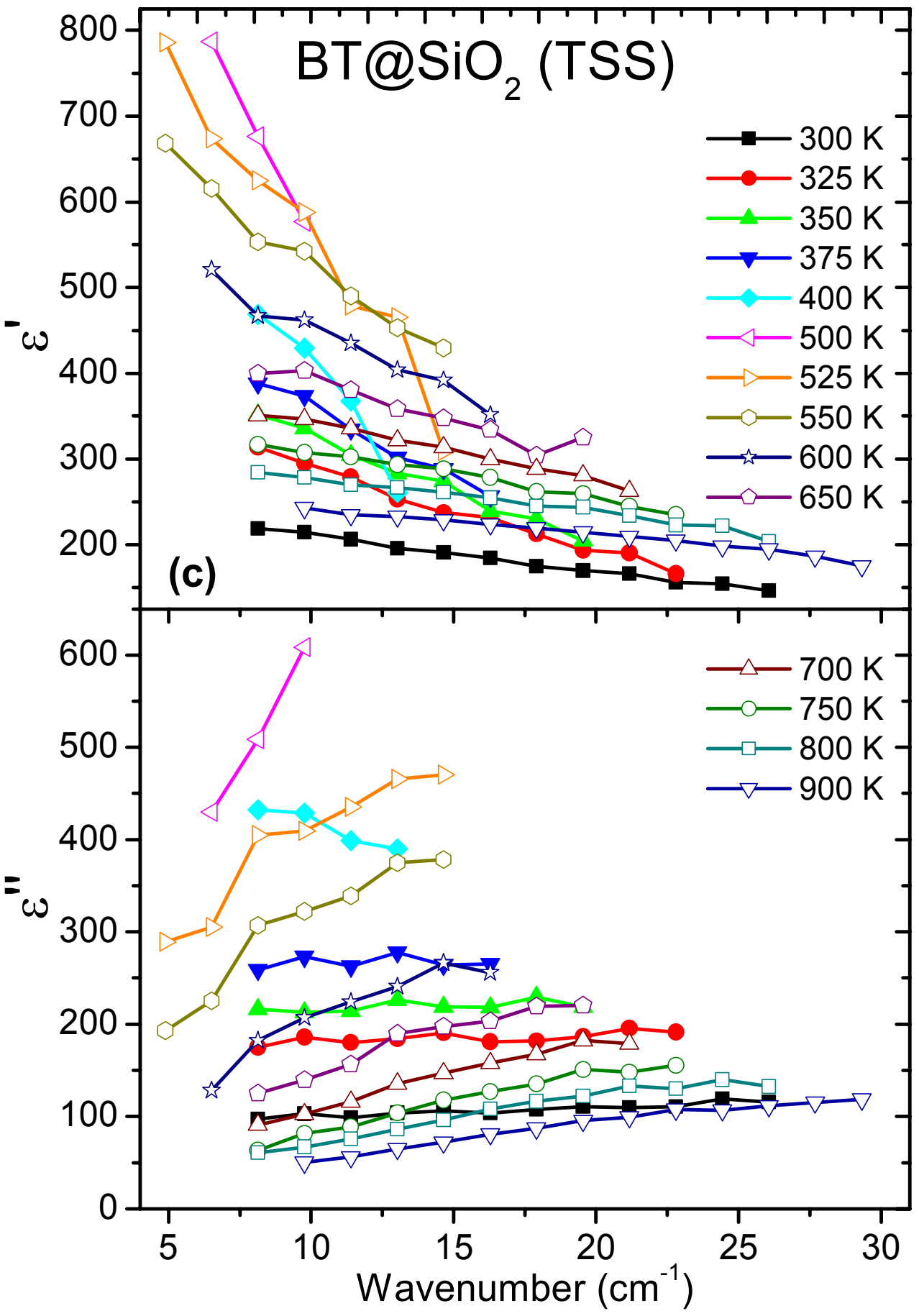

Fig. 3(c).

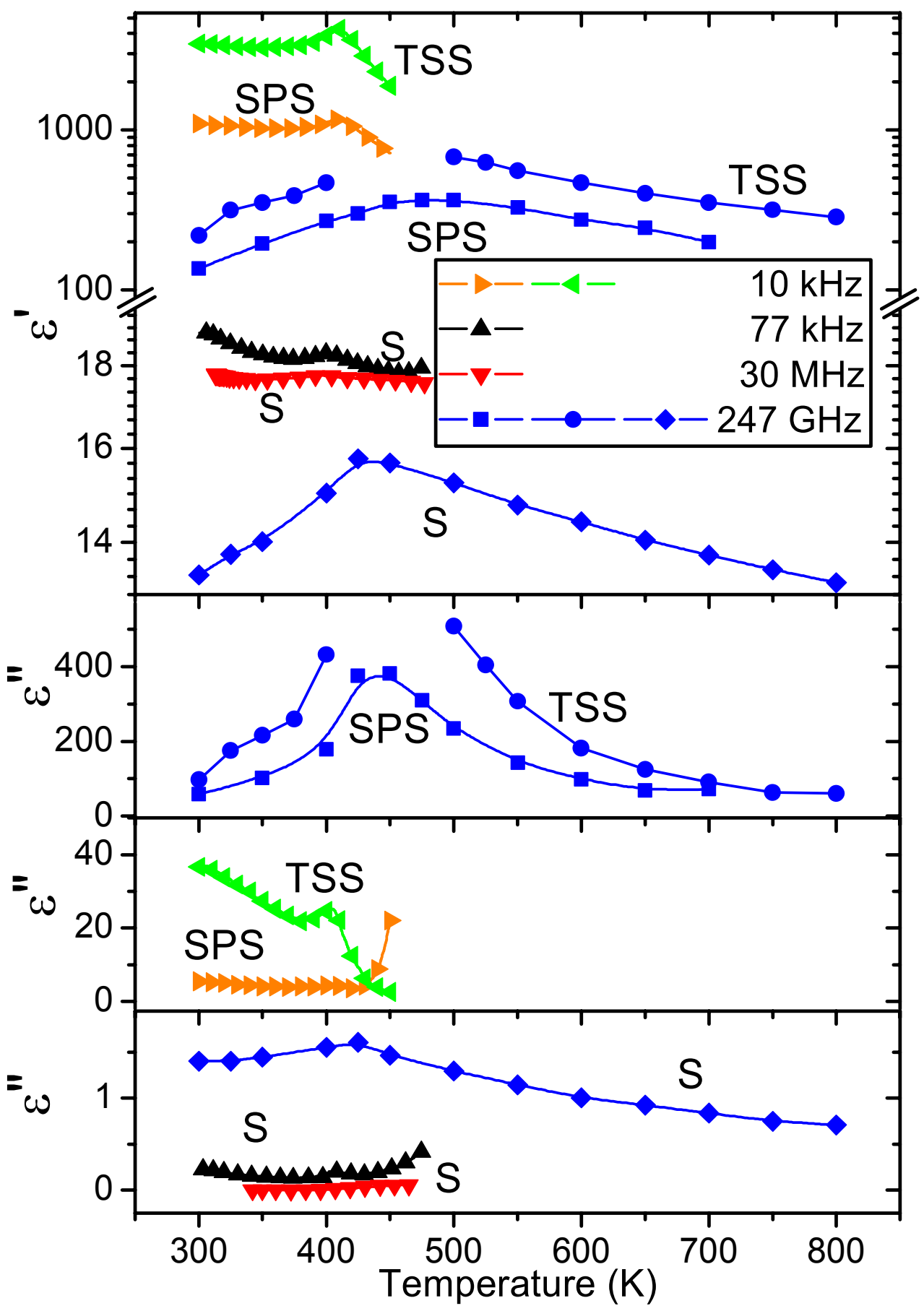

Fig. 4.

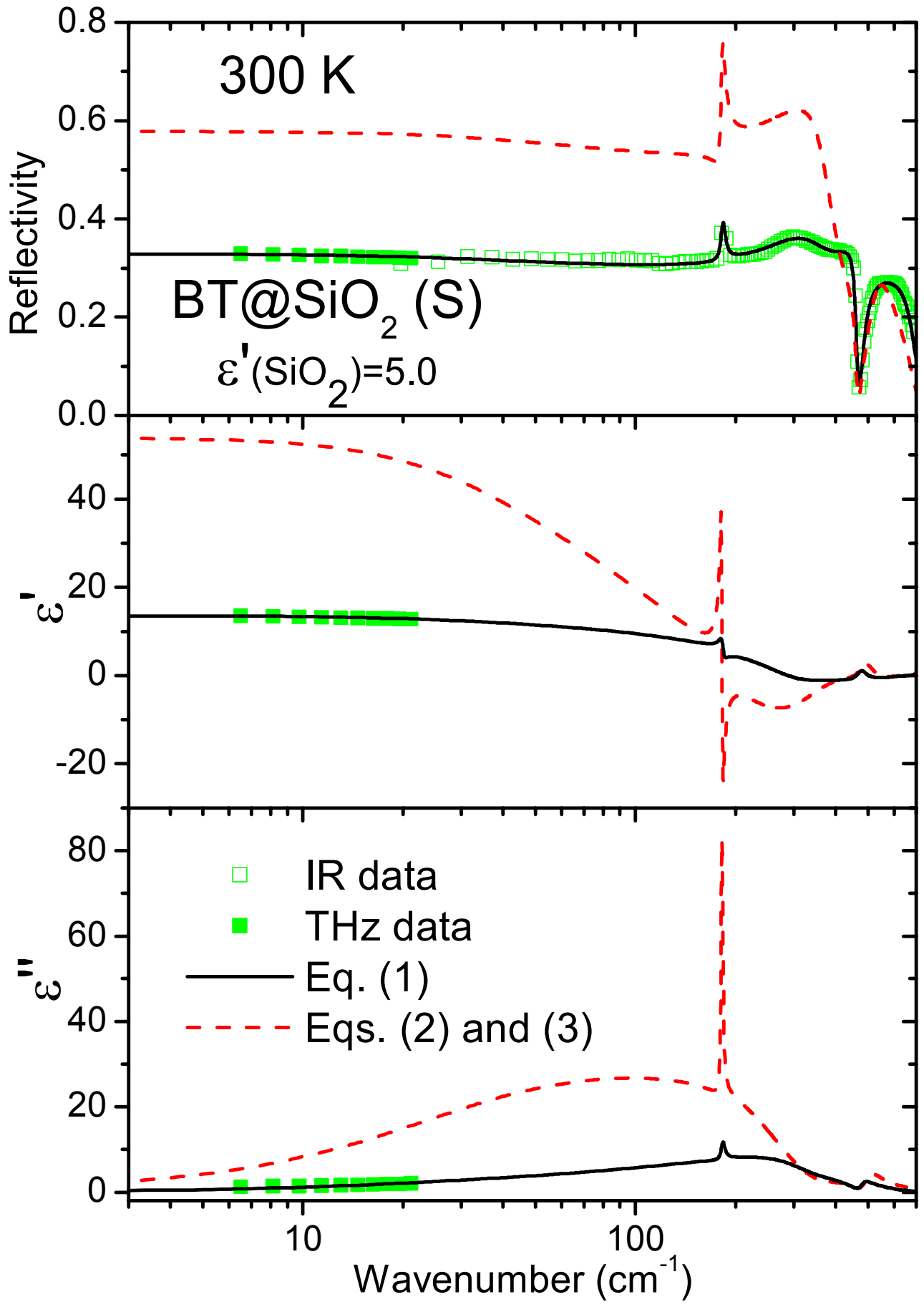

Fig. 5.

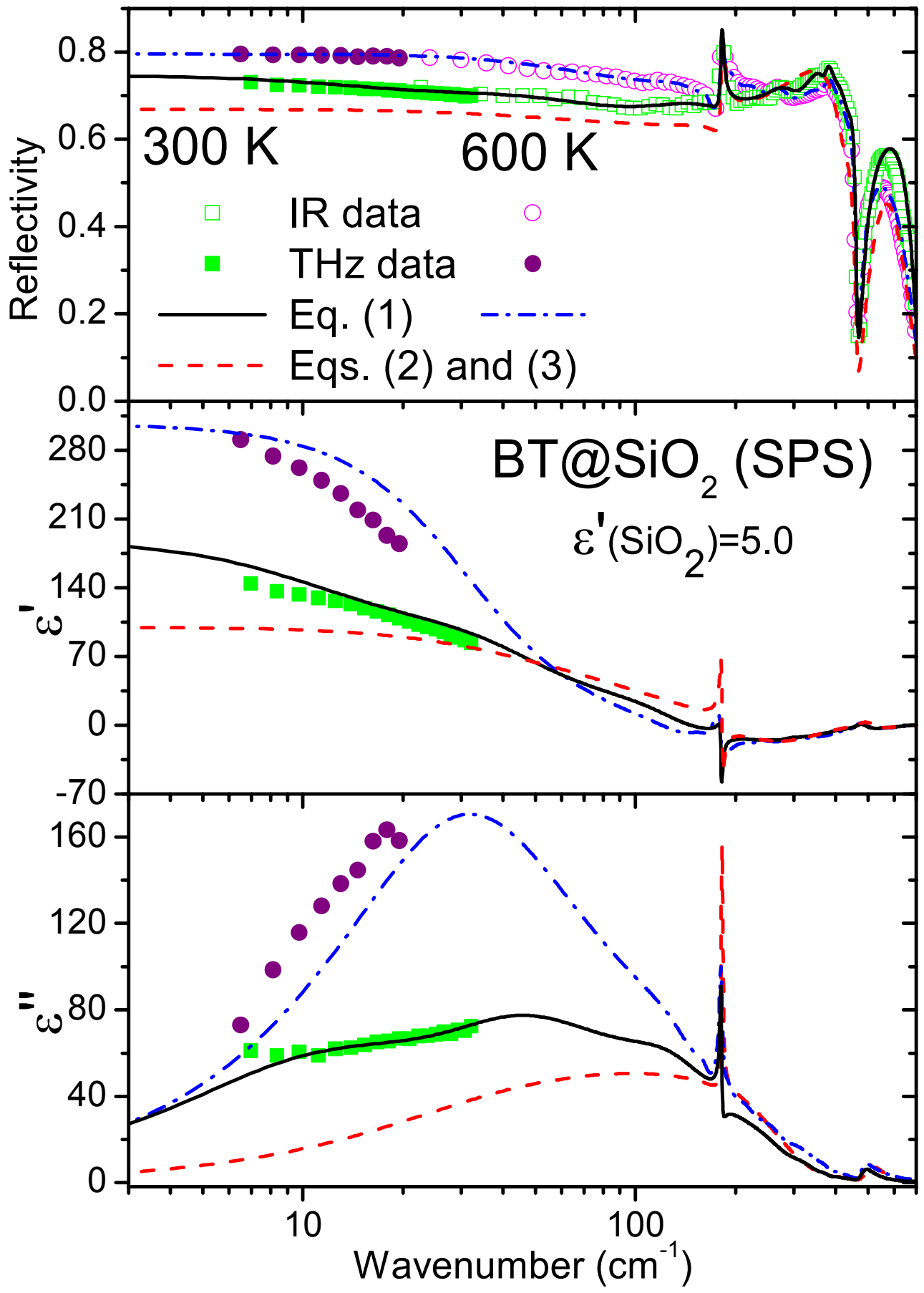

Fig. 6.

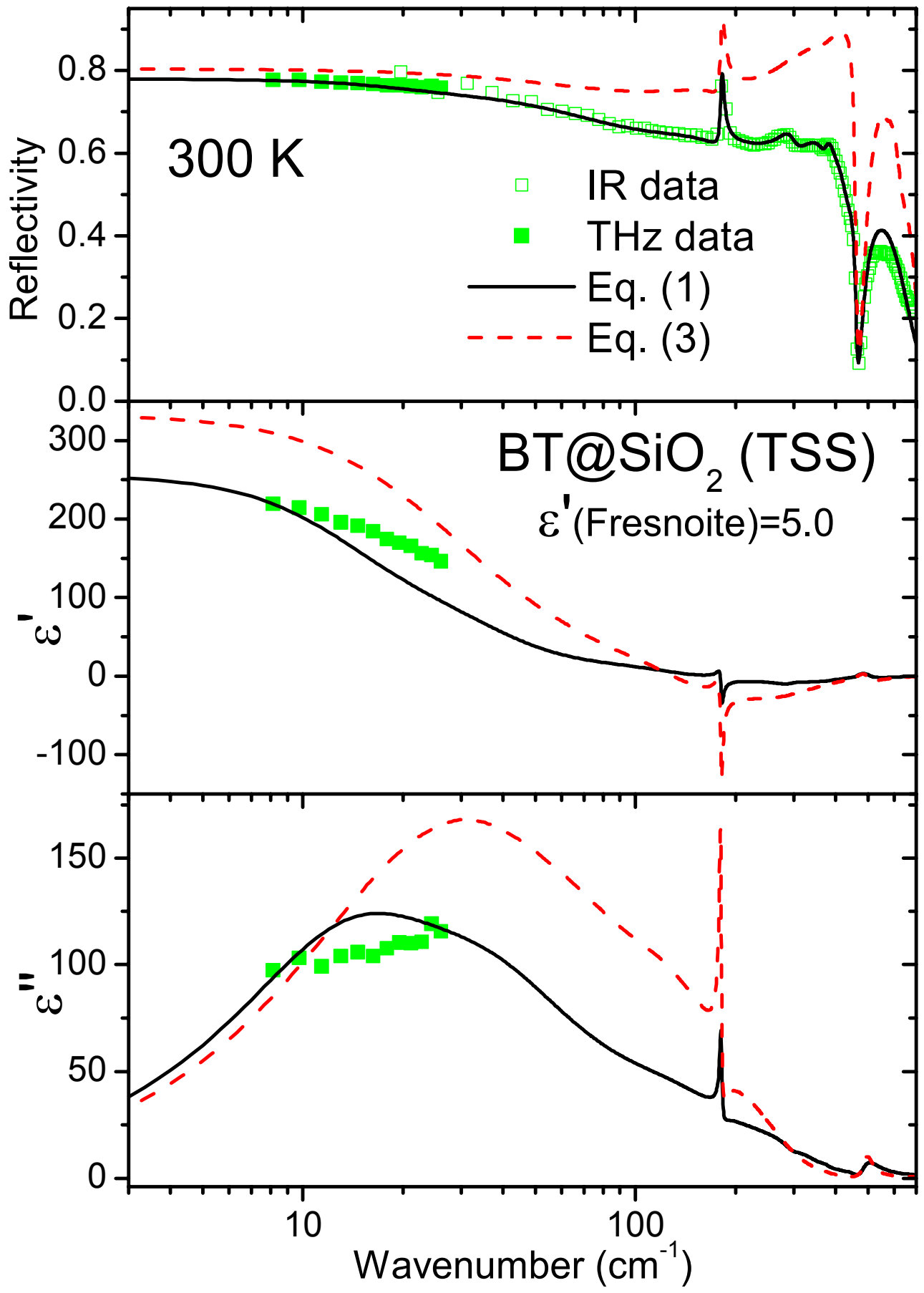

Fig. 7.

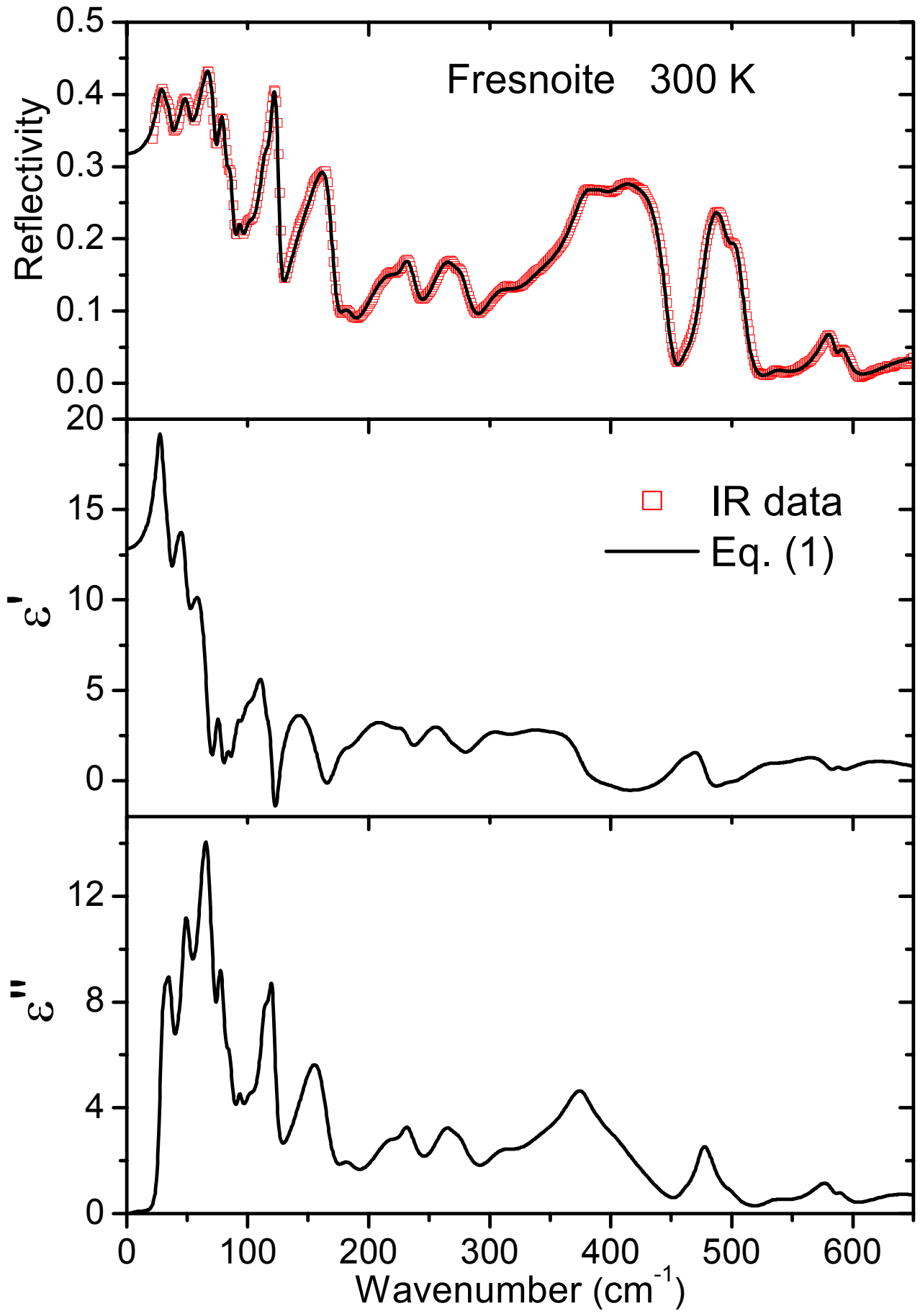

Fig. 8.